\documentclass[aps,prl,twocolumn,showpacs,superscriptaddress,floatfix]{revtex4}

\usepackage{graphicx}
\usepackage{epsfig}
\usepackage[english]{babel}

\newtheorem{lemma}{Lemma}

\newcommand{\be}{\begin{equation}}
\newcommand{\bea}{\begin{eqnarray}}
\newcommand{\eea}{\end{eqnarray}}
\newcommand{\ee}{\end{equation}}

\def\one{\ensuremath{\hbox{$\mathrm I$\kern-.6em$\mathrm 1$}}}

\def\qed{\leavevmode\unskip\penalty9999 \hbox{}\nobreak\hfill
     \quad\hbox{\leavevmode  \hbox to.77778em{%
               \hfil\vrule   \vbox to.675em%
               {\hrule width.6em\vfil\hrule}\vrule\hfil}}
     \par\vskip3pt}
\newcommand{\beaa}{\begin{eqnarray*}}
\newcommand{\eeaa}{\end{eqnarray*}}
\newcommand{\bma}{\begin{subequations}}
\newcommand{\ema}{\end{subequations}}

\def\one{{\bf 1}}

\begin{document}

\title{Matrix product states represent ground states faithfully}
\author{F. \surname{Verstraete}}
\affiliation{Institute for Quantum Information, Caltech, Pasadena, US.}
\author{J. I. \surname{Cirac}}
\affiliation{Max-Planck-Institut f\"ur Quantenoptik, Hans-Kopfermann-Str. 1,
  Garching, D-85748, Germany.}


\begin{abstract}
We quantify how well matrix product states approximate exact ground states of 1-D quantum spin systems as a
function of the number of spins and the entropy of blocks of spins. We also investigate the convex set of local
reduced density operators of translational invariant systems. The results give a theoretical justification for
the high accuracy of renormalization group algorithms, and justifies their use even in the case of critical
systems. \end{abstract}


\maketitle

\section{Introduction}

Quantum spin systems show a very rich variety of fascinating phenomena. Despite the difficulty of describing such
systems due to the large number of degrees of freedom, the development of both analytical and numerical
techniques have allowed us to investigate and understand many of those phenomena. This success story is due, in
part, to the discovery of renormalization group methods \cite{Wilson,DMRG} which are very powerful scalable
numerical techniques which seem to describe amazingly well the ground states of 1D spin systems with local
interactions in terms of matrix product states (MPS) \cite{MPS2,Vwilson}.

On the other hand, one may also investigate many--body quantum systems with a quantum computer or a quantum
simulator \cite{Lloyd}. For instance, one may use an adiabatic algorithm \cite{Fahri} to prepare the quantum
simulator in the ground state of a prescribed Hamiltonian \cite{Murg}, and then measure all one- and two-particle
correlation functions. It seems that this approach may allow us to investigate the properties of many-body
quantum systems that otherwise would not be describable by ordinary computers using existing numerical
algorithms. In fact, an important effort is being made in order to build quantum simulators for some specific
tasks \cite{CZ}.

In this paper we present a series of analytical results which quantify the accuracy of the renormalization group
methods. First, we give an expression that bounds the errors made by approximating general states of $N$ spins by
MPS in terms of the Renyi entropies corresponding to different subsystems. This result, when combined with the
scaling of those Renyi entropies for critical 1D systems, implies that both short- and long-range properties of
the low energy states of critical spin chains can be very precisely described in terms of MPS with a number of
parameters which only scales polynomically with $N$. This indicates that for those systems a quantum computer or
simulator may not give us a big advantage with respect to a classical one. Second, we determine a bound on the
error made in the approximation of 2--particle reduced density operators coming from translationally invariant
states in 1D by MPS in the limit $N\to \infty$. Finally, we illustrate the accuracy of renormalization group
algorithms by some examples.

\section{Ground states as convex problems}

The determination of the ground state of a quantum system consisting of interacting spins on a lattice is highly
nontrivial due to frustration effects. If one considers e.g. a Heisenberg antiferromagnetic interaction between
neighboring spins on a 1-D spin chain, the energy would be minimized if all the reduced density operators of all
neighboring spins would correspond to singlets. The monogamy properties of entanglement
\cite{Werner,Terhal,Wolf,Doherty} however imply frustration effects which forbid the existence of such a state:
if a particle $A$ is maximally entangled with $B$, then it cannot be entangled with $C$. From the technical point
of view, this can readily be proven by showing that the convex set of states obeying the semidefinite constraints
$\rho_{ABC}\geq 0;\rm{Tr}_C\left(\rho_{ABC}\right)=|S\rangle\langle S|=\rm{Tr}_B\left(\rho_{ABC}\right)$ with
$|S\rangle=(|01\rangle-|10\rangle)/\sqrt{2}$ is empty. The frustration effects become stronger when the
coordination number of the lattice increases \cite{Werner,Doherty}. By invoking the quantum de-Finetti theorem
one can prove that in the limit of large coordination number all possible ground states will be separable
\cite{Werner,Werner2}, which on its turn implies that mean-field theory becomes exact in the limit of infinite
dimensional lattices.

Due to the variational nature of ground states, there always exists a ground state with the same symmetries as
the associated Hamiltonian. If the Hamiltonian has translational symmetry and consists of 2-body nearest neighbor
interactions, then it is clear that the energy of a state with the right symmetry is completely determined by its
reduced density operator of 2 neighboring spins. The reduced density operators arising from these (eventually
mixed) states with a given symmetry form a convex set, and the energy for a given Hamiltonian will be minimized
for a state whose reduced density operator is an extreme point in this set. More specifically, the equation
$\rm{Tr}(H\rho)=E$ determines a hyperplane in the space of reduced density operators of states with a given
symmetry, and the energy will be extremal when the hyperplane is tangent to the convex set ($E=E_{extr}$). The
problem of finding the ground state energy of nearest neighbor translational invariant Hamiltonians is therefore
equivalent to the determination of the convex set of 2-body reduced density operators arising from states with
the right symmetry. Strictly speaking, these two problems are dual to each other. In the case of quadratic
Hamiltonians involving continuous variables, the determination of this convex set was solved for fairly general
settings in \cite{Wolf0} by means of Gaussian states. The determination of this convex set in the case of spin
systems however turns out to be much more challenging.

Let us illustrate this with a simple example. Consider the XXZ-Hamiltonian with nearest neighbor interactions
\[\mathcal{H}=-\sum_{<i,j>}S^x_i S^x_j+S^y_iS^y_j+\Delta S^z_iS^z_j\]
on a lattice of arbitrary geometry and dimension\footnote{We assume that the graph corresponding to the lattice
is edge-transitive, meaning that any vertex can be mapped to any other vertex by application of the symmetry
group of the graph.}. Due to the symmetries, the reduced density operator of two nearest neighbors can be
parameterized by only two parameters\footnote{This can easily be proven by invoking local twirling operations
which leave the Hamiltonian invariant.}:
\[\rho=
\frac{1}{4}\left(\openone\otimes\openone+x (\sigma_x\otimes
\sigma_x+\sigma_y\otimes\sigma_y)+z\sigma_z\otimes\sigma_z\right).\] Positivity of $\rho$ enforces $-1\leq z\leq
1-2|x|$, and the state is separable iff $1+z\leq 2|x|$. In the case of an infinite 1-D spin chain, the ground
state energy $E(\Delta)$ has been calculated exactly \cite{Yang}, and this determines the tangent hyperplanes
\[2x+z\Delta+E(\Delta)=0\] whose envelope makes up the extreme points of the convex set of reduced density
operators of translationally invariant 1-D states: the boundary of this convex set is parameterized by
\begin{eqnarray*}z&=&-\partial E(\Delta)/\partial\Delta\\x&=&-(E(\Delta)+\partial E(\Delta)/\partial\Delta)/2,\end{eqnarray*}
which we plotted in Figure 1. We also plot the boundary for the 2-dimensional square lattice. These 2-D data were
obtained by numerical methods \cite{2D,Suljavson,Lin}); of course this convex set is contained in the previous
one, as all the semidefinite constraints defining the set corresponding to 1-D are strictly included in the set
of constraints for the 2-D case. Finally, we plot the set of separable states, which contains the reduced density
operators of the allowed states for a lattice with infinite coordination number. The boundary of this separable
set is given by the inner diamond; this immediately implies that the difference between the exact energy and the
one obtained by mean field theory will be maximized whenever the hyperplane that forms the boundary of the first
set will be parallel to this line. This happens when $\Delta=-1$ (independent of the dimension!), which
corresponds to the antiferromagnetic case, and this proves that the "entanglement gap" \cite{Doherty} in the
XXZ-plane is maximized by the antiferromagnetic ground state for any dimension and geometry. Similarly, it proves
that the ground state is separable whenever $\Delta\geq 1$ and $\Delta=-\infty$. Note also that in the 2-D case,
part of the boundary of the convex set consists of a plane parameterized by $2x+z+E(1)=0$. This indicates a
degeneracy of the ground state around the antiferromagnetic point, and indicates that a phase transition is
occurring at that point (more specifically between an Ising and a Berezinskii- Kosterlitz-Thouless phase). It
would be very interesting to investigate this further, and it is fascinating that this phase transition could be
detected by just looking at the structure of these low-dimensional convex sets.

\begin{figure}[t]
  \includegraphics[width=\linewidth]{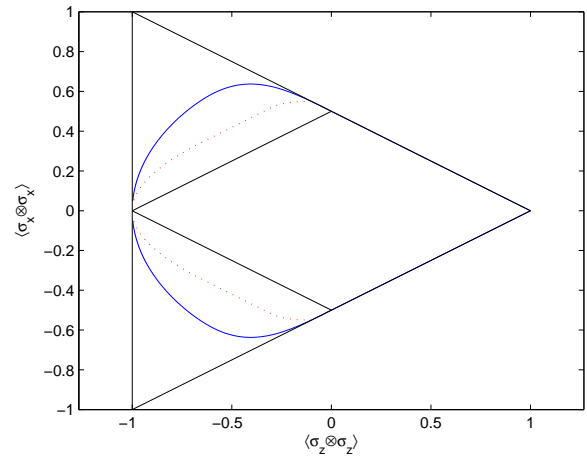}
  \caption{Convex sets of the possible reduced density operators of translational invariant states in the XX-ZZ plane: the big triangle
  represents all positive density operators; the inner
  parallellogram represents the separable states; the union of the separable cone and the convex hull of the full
  curved line is the complete convex set in the case of a 1-D
  geometry, and the dashed lines represent extreme points in the 2-D case of a square lattice. The singlet corresponds to the point with coordinates (-1,-1).}
  \label{convex}
\end{figure}

The previous discussion implies that ground states of quantum spin systems are very special: they are completely
determined by their 2-body reduced density operators. Typically, it even holds that the ground state of an
interacting spin system is unique, which implies that most extreme points in the convex set of reduced density
operators uniquely characterize a state with the right symmetry properties. This is very good news if we want to
create families of variational ground states: it suffices to approximate well the local properties of all
translational invariant states. The family of matrix product states (MPS) \cite{Fannes,MPS2,VPC04} and
generalizations to higher dimensions (PEPS) \cite{2D,2Dd,Nishino} were exactly created with this property in
mind; the amazing accuracy of renormalization group algorithms is precisely related to the fact that the convex
set under consideration can be very well approximated with the reduced density operators of MPS. Both Wilson's
numerical renormalization group \cite{Wilson} and DMRG \cite{DMRG,DMRG1} can indeed be reformulated as
variational methods within the MPS \cite{VPC04,Vwilson}.

\begin{figure}[t]
  \includegraphics[width=\linewidth]{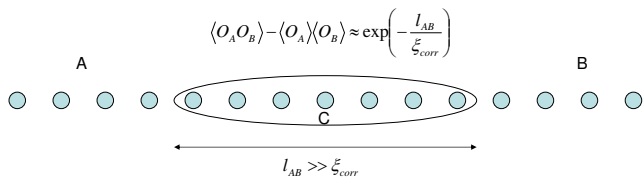}
  \caption{A one dimensional spin chain with finite correlation length $\xi_{corr}$; $l_{AB}$ denotes the distance between the block $A$ (left) and $B$ (right). Because $\l_{AB}$ is much larger than the correlation length $\xi_{corr}$, the state $\rho_{AB}$ is essentially a product state.}
  \label{figurecorr}
\end{figure}

The reason why ground states of gapped quantum spin systems can be parameterized as MPS/PEPS can be understood by
the following handwaving argument. A celebrated theorem of Hastings \cite{Hastings} states that in the case of a
gapped system all correlations are decaying exponentially. Let us therefore consider a 1-D gapped quantum spin
system with correlation length $\xi_{corr}$. Due to the finite correlation length, it is expected that the
reduced density operator $\rho_{AB}$ obtained when tracing out a block $C$ of length $l_{AB}\gg \xi_{corr}$ (see
figure \ref{figurecorr}) is equal to \begin{equation} \rho_{AB}\simeq
\rho_A\otimes\rho_B\label{hasg}\end{equation} up to exponentially small corrections. As noted by T. Osborne,
Hastings theorem does not imply the validity of equation (\ref{hasg}), as it was shown  in \cite{datahiding} that
orthogonal states exist whose correlation functions are exponentially close to each other; although it would be
very surprising that ground states would exhibit that property, this prohibits to turn the presented handwaving
argument into a rigorous one. But let's for the sake of the argument assume that equation (\ref{hasg}) is true.
Then, the original ground state $|\psi_{ABC}\rangle$ is a purification of this mixed state, but it is of course
also possible to find a purification of the form $|\psi_{AC_l}\rangle\otimes|\psi_{BC_r}\rangle$ (up to
exponentially small corrections) with no correlations whatsoever between $A$ and $B$; here $C_l$ and $C_r$
together span the original block $C$. It is however well known that all possible purifications of a mixed state
are equivalent to each other up to local unitaries on the ancillary Hilbert space. This automatically implies
that there exists a unitary operation $U_C$ on the block $C$ (see figure \ref{figurecorr}) that completely
disentangles the left from the right part:
\[\openone_A\otimes U_{C}\otimes\openone_C|\psi_{ABC}\simeq|\psi_{AC_l}\rangle\otimes|\psi_{BC_r}\rangle.\]
This implies that there exists a tensor $A_{\alpha,\beta}^i$ with indices $1\leq \alpha,\beta,i\leq D$ (where $D$
is the dimension of the Hilbert space of $C$) and states
$|\psi^A_\alpha\rangle,|\psi_i^C\rangle,|\psi^B_\beta\rangle$ defined on the Hilbert spaces belonging to $A,B,C$
such that
\[|\psi_{ABC}\rangle\simeq \sum_{\alpha,\beta,i}A^i_{\alpha,\beta}|\psi^A_\alpha\rangle|\psi^C_i\rangle|\psi^B_\beta\rangle.\]
Applying this argument recursively leads to a MPS (to be defined in equation (\ref{MPS})) and indeed gives a
strong hint that ground states of gapped Hamiltonians are well represented by MPS. In this paper we will prove
that this is indeed true, even in the case of critical systems.

As a good illustration of the actual accuracy obtained with MPS, we calculated the convex set obtained with MPS
in the thermodynamic limit for the XXZ-chain with $D=2$, where $D$ is the dimension of the matrices in the MPS.
It is almost unbelievable how good the exact convex set can be approximated. Note that typical DMRG calculations
have $D\sim 200$, and that the accuracy grows superpolynomial in $D$. Note also that the $D=1$ case corresponds
to mean-field theory, whose corresponding convex set coincides with the set of separable states.

\begin{figure}[t]
  \includegraphics[width=.95\linewidth]{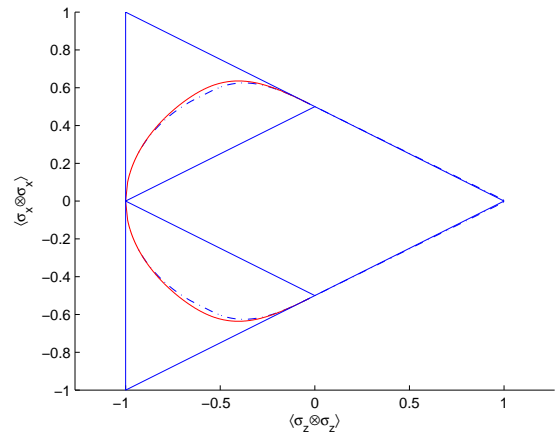}
  \caption{Convex sets in the XXZ-plane: the inner diamond borders the set
  of separable states (see Fig. 1). Dash-dotted:
  extreme points of the convex set produced by MPS of $D=2$.}
  \label{error}
\end{figure}

The same argument involving the notion of a correlation length applies in higher dimensions and indicates that
PEPS represent ground states of gapped local Hamiltonians well. Note however that the convex set in the 2-D case
is much closer to the separable one than in the 1-D case; this gives a hint that PEPS of smaller dimension will
suffice to give the same accuracy as in the 1-D case. In the next section we will quantitatively bound how well a
translationally invariant state can be represented in terms of a MPS, and will analyze the corresponding
implications for the description of ground states of 1D spin chains.

\section{Analytical bounds for matrix product states}

Let us start by recalling the expression of $D$--dimensional MPS describing $N$ spins of dimension $d$,
 \be
 \label{MPS}
 |\psi_D\rangle = \sum_{i_1,\ldots,i_N=1}^d
 {\rm tr}\left[A^{[1]i_1} \ldots A^{[N]i_N}\right] |i_1,\ldots,i_N\rangle.
 \ee
Here, $A^{[k],1},\ldots,A^{[k],d}$ are, in general, $D_k\times \tilde D_k$ complex matrices, with $D_{k+1}=\tilde
D_k\le D$.

We will show first how well one can describe general states in terms of MPS. We will choose $D_1=\tilde D_N=1$,
i.e., $A^{[1]i_1}$ and $A^{[N]i_N}$ will be vectors. In this case, one can impose the following \emph{gauge}
condition \cite{MPSrenorm,VidalMPS}:
 \be
 \label{gauge}
 \sum_i A^{[m]i}A^{[m]i\dagger}=\openone,\quad
 \sum_i A^{[m]i\dagger}\Lambda^{[m+1]}A^{[m]i}=\Lambda^{[m]}.
 \ee
Here $\Lambda^{[m]}$ represents the diagonal matrix with the corresponding eigenvalues $\{\tilde\lambda^{[m]i}\}$
sorted in decreasing order, i.e. $\tilde\lambda^{[m]1}\ge \tilde\lambda^{[m]2}\ge \ldots$. We consider an
arbitrary state $|\psi\rangle$ and denote by \[\{\mu^{[\alpha]i}\}, i=1..N_\alpha=d^{\min(\alpha,N-\alpha)}\] the
eigenvalues of the reduced density operators
\[\rho_\alpha=\rm{Tr}_{\alpha+1,\alpha+2,\ldots
,N}|\psi\rangle\langle\psi|,\] also sorted in decreasing order.

In the following we will investigate: (i) How well a general quantum state can be approximated by a
$D$--dimensional MPS; (ii) How well the reduced density operators of a translationally invariant state can be
describe by the one corresponding to a $D$--dimensional MPS.

\subsection{General states: analytic results and implications}

We start out with the main result of this paper, which gives an upper bound to the error made by approximating a
general state by a MPS. As we will show below, this has very important implications in the performance of the
renormalization algorithms to describe ground states of 1D spin chains.

\begin{lemma}
There exists a MPS $|\psi_D\rangle$ of dimension $D$ such that
 \[\||\psi\rangle-|\psi_D\rangle\|^2\leq
2\sum_{\alpha=1}^{N-1}\epsilon_\alpha(D)
\]
where $\epsilon_\alpha(D)=\sum_{i=D+1}^{N_\alpha}\mu^{[\alpha]i}$.
\end{lemma}

\emph{Proof:} We can always write $\psi$ as a MPS of dimension $D=2^{N/2}$ and fulfilling (\ref{gauge}). Let us
now consider the $D$-dimensional MPS $|\psi_D\rangle$ which is defined by the $D\times D$ matrices
$[A^{[\alpha]i}]_{(1\ldots D,1\ldots D)}$ (i.e. the upper--left block of $A^{[\alpha]i}$). The goal is now to
bound $\langle\psi|\psi_D\rangle$. The gauge conditions were chosen such as to make the task simple:
 \be
 \langle\psi_D|\psi\rangle={\rm Tr}\left[\$_2\left(\cdots
 \$_{N-2}\left(\$_{N-1}\left(\Lambda^{N-1}P\right)P\right)
 P\cdots\right)P\right];\label{HJD}
 \ee
here $P=\sum_{k=1}^D|k\rangle\langle k|$ and $\$_m(X)=\sum_i A^{[m]i\dagger}XA^{[m]i}$ represents a
trace-preserving completely positive map (TPCP-map) parameterized by the Kraus operators $A^{[m]i}$. Let us now
recursively define
\[Y^{[k]}=\$_{k}\left(Y^{[k+1]}P\right),\hspace{.5cm}
Y^{[N-1]}=\Lambda^{[N-1]}P;\] observe that $\Lambda^{[k]}=\$_{k}\left(\Lambda^{[k+1]}\right)$. We want a bound on
$\rm{Tr}|\Lambda^{[1]}-Y_1|$, as equation (\ref{HJD}) is equal to $\rm{Tr}(Y^{[2]})$. The crucial property we
need is that TPCP-maps are contractive with relation to the trace-norm \footnote{This can directly be proven by
considering the Neumark representation of a TPCP-map as a unitary in a bigger space.}: $\rm{Tr}|\$(X)|\leq
\rm{Tr}|X|$. It follows that
\begin{eqnarray*} &&\rm{Tr}|\Lambda^{[k]}-Y^{[k]}|=\rm{Tr}|\$_k\left(\Lambda^{[k+1]}-Y^{[k+1]}P\right)|\leq\\&&\leq
\rm{Tr}|\Lambda^{[k+1]}-Y^{[k+1]}P|\\&&
\leq\rm{Tr}|\Lambda^{[k+1]}-Y^{[k+1]}|+\rm{Tr}|\Lambda^{[k+1]}(\openone-P)|.\end{eqnarray*} Note that the last
term in the sum is exactly given by $\sum_{\alpha=D+1}^{2^{N/2}}\lambda^{[k+1]\alpha}$. The theorem now follows
immediately by recursion and by observing that $\langle\psi_D|\psi_D\rangle\leq 1$ by similar arguments.\qed

The implications of this result are very strong: it shows that for systems for which the $\epsilon_\alpha(D)$
decay fast in $D$, there exist MPS with \emph{small} $D$ which will not only reproduce well the local
correlations (such as energy) but also all the nonlocal properties (such as correlation length). The following
lemma now relates the derived bound to the Renyi entropies of the reduced density operators, through which one
can make the connection to the ground states of 1D Hamiltonians. The Renyi entropies of $\rho$ are defined as
\[S^\alpha(\rho)=\frac{1}{1-\alpha}\log\left({\rm Tr}\rho^\alpha\right),\] and we will consider  $0<\alpha< 1$. We
denote as before $\epsilon(D)=\sum_{i=D+1}^\infty\lambda_i$ with $\lambda_i$ the nonincreasingly ordered
eigenvalues of $\rho$. Then we have

\begin{lemma}
Given a density operator $\rho$. If $0<\alpha<1$, then $\log(\epsilon(D))\leq
\frac{1-\alpha}{\alpha}\left(S^\alpha(\rho)-\log\frac{D}{1-\alpha}\right)$.
\end{lemma}

\emph{Proof:} Let us first characterize the probability distribution that has maximal possible weight in its tail
(i.e. $p=\sum_{i=D+1}^\infty p_i$) for a given Renyi-entropy. Introducing a free parameter $0<h\leq (1-p)/D$,
such a probability distribution must be of the form
\begin{eqnarray*}p_1&=&1-p-(D-1)h\\ h&=&p_2=p_3=\cdots p_{D+p/h}\\
p_{D+p/h+1},\cdots p_{\infty}&=&0\end{eqnarray*} because this distribution majorizes all other ones with given
$p,D,p_D$ (Renyi-entropies are Schur-convex functions). For a given $p,D,h$, it holds that
\begin{eqnarray*}\sum_i p_i^\alpha&=&(1-p-(D-1)h)^\alpha+(D-1+p/h)h^\alpha\\&\geq& Dh^\alpha+ph^{\alpha-1}.\end{eqnarray*} Minimizing this
expression with relation to $h$, we get \[\sum_i p_i^\alpha\geq
(D^{1-\alpha}p^\alpha)/((1-\alpha)^{1-\alpha}\alpha^\alpha).\] Denoting $S^\alpha(p,D)$ the minimal possible
entropy for given $p,D$, we get \[S^\alpha(p,D)\geq \frac{1}{1-\alpha}\log\left(
\frac{D^{1-\alpha}p^\alpha}{(1-\alpha)^{1-\alpha}\alpha^\alpha}\right)\] and hence \[p\leq
\exp\left(\frac{1-\alpha}{\alpha}\left(S^\alpha(p,D)-\log\frac{D}{1-\alpha}\right)\right).\] The proof now
follows by replacing $S^\alpha(p,D)$ by $S^\alpha(\rho)$. \qed

This lemma is very interesting in the light of the fact that in the case of critical systems, arguable the
hardest ones to simulate \footnote{For non--critical systems, the renormalization group flow is expected to
increase the Renyi entropies in the UV direction. The corresponding fixed point corresponds to a critical system
whose entropy thus upper bounds that of the non--critical one.}, the Renyi-entropy of a contiguous block of $L$
spins scales as \cite{Peshel,Kitaev,Korepin,Cardy,Vidal}
 \be
 S^\alpha(\rho_L)\simeq\frac{c+\bar{c}}{12}
 \left(1+\frac{1}{\alpha}\right)\log(L)\label{infg}
 \ee
for all $\alpha>0$; here  $c$ is the central charge. The fact that the eigenvalues of $\rho_L$ decay fast has
previously been identified as a indication for the validity of the DMRG-approach \cite{Peshel}. The truncation
error \cite{DMRG,DMRG1}, which has been used in the DMRG community as a check for convergence, is essentially
given by $\epsilon(D)-\epsilon(2D)$ and therefore indeed gives a good idea of the error in a simulation.

Let us investigate how the computational effort to simulate such critical systems scales as a function of the
length $N=2L$ of the chain. Let us therefore consider the Hamiltonian associated to a critical system, but
restrict it to $2L$ sites. The entropy of a half chain (we consider the ground state $|\psi_{ex}\rangle$ of the
finite system) will typically scale as in eq. (\ref{infg}) but with an extra term that scales like $1/N$. Suppose
we want to enforce that $\||\psi_{ex}\rangle-|\psi_D\rangle\|^2\leq\epsilon_0/L$ with $\epsilon_0$ independent of
$L$ \footnote{We choose the $1/L$ dependence such as to assure that the absolute error in extensive observables
does not grow.}. Denote the minimal $D$ needed to get this precision for a chain of length $2L$ by $D_L$.
Following lemma (1) and the fact that the entropy of all possible contiguous blocks reaches its maximum in the
middle of the chain (hence $p\leq\epsilon_0/L^2$ is certainly sufficient), lemma (1) and (2) combined yield
\[D_L\leq
cst\left(\frac{L^2}{(1-\alpha)\epsilon_0}\right)^{\frac{\alpha}{1-\alpha}}
L^{\frac{c+\bar{c}}{12}\frac{1+\alpha}{\alpha}}.\] This shows that $D$ only has to scale polynomially in $L$ to
keep the accuracy $\epsilon_0/L$ fixed; in other words, there exists an efficient scalable representation for
ground states of critical systems (and hence also of noncritical systems) in terms of MPS! Such a strong result
could not have been anticipated from just doing simulations.

Now what about the complexity of finding this optimal MPS? It has been observed that DMRG converges exponentially
fast to the ground state with a relaxation time proportional to the inverse of the gap $\Delta$ of the system
\cite{DMRG1}. For translational invariant critical systems, this gap seems to close only polynomially. As we have
proven that $D$ only have to scale polynomially too, the computational effort for finding ground states of 1-D
quantum systems is polynomial ($P$). Let us recapitulate for which systems and under which conditions this
statement is true: 1) the $\alpha$-entropy of blocks in the exact ground state grow at most logarithmically with
the size of the block for some $\alpha<1$; 2) the gap of the system scales at most polynomially with the system
size; 3) given a gap that obeys condition 2, there exists an efficient DMRG-like algorithm that converges to the
global minimum. As the variational MPS approach \cite{VPC04} is essentially an alternating least squares method
of solving a non-convex problem, there is a priori no guarantee that it will converge to the global optimum
\footnote{Let us e.g. specify a method which should in principle not get trapped in local minima. Like in the
adiabatic theorem, we can construct a time dependent Hamiltonian $H(t)$ with $H(0)$ trivial and $H(1)$ the
Hamiltonian to simulate; if we discretize this evolution in a number of steps that grows polynomially in the
inverse gap, the adiabatic theorem guarantees that we will end up in the ground state of $H(1)$ if we can follow
the ground state of $H(t)$ closely. The idea is to make $D$ of $|\psi_D(t)\rangle$ large enough such as to follow
the ground state $|\psi(t)\rangle$ close enough in such a way that the optimization is always convex around the
global optimum within the domain $\|\chi\rangle-|\psi(t)\rangle\|\leq\epsilon$.}, although the occurrence of
local minima seems to be very unlikely \cite{DMRG1}. Note that this e.g. implies that there would be no
exponential gain in using a quantum computer in simulating 1-D spin systems, as the steps needed in a quantum
computer are also bounded by the inverse of the gap!

\subsection{Reduced density operators of translationally invariant
states}

Now we turn to the study of how MPS approximate the local properties of translationally invariant states $\psi$
of $N\to \infty$ systems. Let us denote by $\psi$ a translationally invariant state with $N=\infty$ and by
$\sigma={\rm tr}(\rho)$ the corresponding two--particle reduced density operator, where the trace is taken with
respect to all particles except for two neighbors (say particle $1$ and $2$). Our goal is to show that there
exists a $D$--dimensional MPS whose corresponding reduced density operator, $\sigma_D$, approximates well
$\sigma$. More specifically, we want to bound $||\sigma-\sigma_D||_1$ from above. Note that any nearest neighbor
correlation function, and in particular the energy density will be automatically bounded.

One can represent the state $\psi$ \cite{Fannescompl} as a MPS (\ref{MPS}) with all the $A$'s equal and
$D,N=\infty$. Without loss of generality we can again choose the {\em gauge} condition (2) with $A,\Lambda$ site
independent. In fact, the elements $\Lambda_{ii}$ coincide with the eigenvalues $\lambda_i$ of the reduced
density operator corresponding to half a chain. As before, they will play a very important role in the bound that
we derive; in particular, we will need $\epsilon(D)=\sum_{\alpha=D+1}^\infty\lambda_\alpha$ and we define
\[\gamma_D=\left(\sum_{\alpha=1}^D\sqrt{\lambda_\alpha}\right)\sum_{\beta=1}^D\sum_{n=1}^\infty
 \frac{\lambda_{\beta+Dn}}{\sqrt{\lambda_\beta}}\]
which roughly behaves like $\epsilon(D)$, i.e. it becomes small when $D$ is large.

\begin{lemma}
Given $\psi$ and $\sigma$ as defined above, there exists a $D$--dimensional MPS such that
 \be
 \label{bound}
 \frac{1}{d^2}\rm{Tr}|\sigma-\sigma_D|\leq
 2(\sqrt{\epsilon(D)}+\epsilon(D))+\gamma_D
 \ee
\end{lemma}

 \emph{Proof:} One can easily show that
 \begin{eqnarray*}
 r^{i,i',j,j'}&\equiv&\langle i|\langle
 i'|\rho|j\rangle|j'\rangle={\rm Tr}\left[A^{j\dagger}\Lambda
 A^{i}A^{i'}A^{j'\dagger}\right]\\
 &=& \sum_{\alpha,\beta,n,m}
 [A^{j\dagger}\Lambda A^i]_{\alpha+Dn,\beta+Dm}
 [A^{i'} A^{j' \dagger}]_{\beta+Dm,\alpha+Dn}
\end{eqnarray*}
Let us consider the $D$-dimensional matrix product density operator (MPDO) \cite{VGC04} \[ \rho_D=\sum{\rm
 Tr}\left(B^{i_1j_1}\ldots B^{i_\infty
 j_\infty}\right)|i_1\rangle\ldots |i_\infty\rangle\langle
 j_1|\ldots\langle j_\infty|\]
parameterized by

\begin{eqnarray*}
B^{ij}_{\alpha\alpha';\beta\beta'}&=&
 \sum_{n,m} A_{\alpha+Dn,\beta+Dm}^i A^{*j}_{\alpha'+Dn,\beta'+Dm}
 x_{\alpha,n}x_{\alpha',n}\\
 x_{\alpha,n}&=&\sqrt{\frac{\lambda_{\alpha+Dn}}{\sum_{m=0}^\infty
 \lambda_{\alpha+Dm}}}
\end{eqnarray*}
  The following eigenvalue equations are easily verified:
\begin{eqnarray*}
 \sum
 B^{ii}_{\alpha\alpha';\beta\beta'}\delta_{\alpha'\beta'}&=&
 \delta_{\alpha\beta}\\\sum
 B^{ii}_{\alpha\alpha';\beta\beta'}\tilde{\Lambda}_{\alpha \beta}&=&
 \tilde{\Lambda}_{\alpha'\beta'}
\end{eqnarray*}
where \[\tilde{\Lambda}_{\alpha\beta}=\delta_{\alpha\beta}\sum_{k=0}^\infty\lambda_{\alpha+Dk}.\] The reduced
density operator $s^{i,i',j,j'}\equiv\langle i|\langle i'|\rho_{D}|j\rangle|j'\rangle$ is given by
 \begin{equation}
 \sum_{\alpha\beta n m}
 \left[A^{j\dagger}\Lambda
 A^{i}\right]_{\alpha+Dn,\beta+Dn}\left[A^{i'}
 A^{j'\dagger}\right]_{\beta+Dm,\alpha+Dm}
 x_{\beta,m}x_{\alpha,m}.\label{EMPDO}
 \end{equation}

We divide $r$ and $s$ in two parts,
 \beaa
 r^{i,i',j,j'} &=& r^{i,i',j,j'}_0 + r^{i,i',j,j'}_1,\\
 s^{i,i',j,j'} &=& s^{i,i',j,j'}_0 + s^{i,i',j,j'}_1.
 \eeaa
The first parts ($r_0$ and $s_0$) contain just the part of the sums with $n=m=0$, whereas the second ones ($r_1$
and $s_1$) contain the rest of the sums.

The goal is to find upper bounds for
 \beaa
 \Delta r &=& \sum_{i,i',j,j'=1}^d |r^{i,i',j,j'}_1|,\\
 \Delta s &=& \sum_{i,i',j,j'=1}^d |s^{i,i',j,j'}_1|,\\
 \Delta t &=& \sum_{i,i',j,j'=1}^d |r^{i,i',j,j'}_0-s^{i,i',j,j'}_0|.
 \eeaa

We start with $\Delta r$. We can write
 \be
 \label{Eqa}
 \Delta r = \sum_{i,i',j,j'=1}^d |{\rm Tr}[A^{j\dagger}\Lambda
 A^i P A^{i'} A^{j'\dagger}]+{\rm Tr}[PA^{j\dagger}\Lambda
 A^i A^{i'} A^{j'\dagger}]|.
 \ee
Here,
 \[
 P=\sum_{x=D}^\infty |x\rangle\langle x|
 \]
 with $|x\rangle$ a unit vector in the computational basis. Using Cauchy--Schwartz inequality we have
 \beaa
&&\sum_{i,i',j,j'=1}^d \left|{\rm Tr}\left[A^{j\dagger}\Lambda
 A^i P A^{i'} A^{j'\dagger}\right]\right| \le\\
&& \hspace{.5cm}\left[ \sum_{i,i',j,j'=1}^d {\rm Tr}[A^{i'\dagger}PA^{i\dagger} \Lambda A^i P A^{i'}]
 \right]^{1/2}\times\\
&& \hspace{1cm} \left[ \sum_{i,i',j,j'=1}^d {\rm Tr}[A^{j'\dagger}A^{j\dagger}
 \Lambda A^j A^{j'}]
 \right]^{1/2}=\\
 &&= d^2 \left[ {\rm Tr}[P\Lambda P]
 \right]^{1/2} = d^2 \sqrt{\sum_{x=D}^\infty \lambda_x}
 \eeaa
The second term in Eq.(\ref{Eqa}) gives the same, so that we have
 \[
 \Delta r \le 2 d^2 \sqrt{\sum_{x=D}^\infty \lambda_x}=2d^2\sqrt{\epsilon(D)}.
 \]

The term $\Delta s$ is a bit more tricky. We write it as
 \[
 \Delta s= \sum_{n,m}' \sum_{x,y} \sum_{i,i',j,j'=1}^d
 L^{ii'}_{x,y,n,m} L^{jj'\ast}_{x,y,n,m},
 \]
where the prime in the sum means that $n=m=0$ is excluded, and
 \[
 L^{ii'}_{x,y,n,m}= \sqrt{\lambda_x} \sum_{\beta} x_{\beta,m}
 A^i_{x,\beta+Dn} A^{i'}_{\beta+Dm,y}.
 \]
Using Cauchy--Schwartz inequality we get
 \begin{eqnarray*}
 \Delta s &\le& \sum_{n,m}' \sum_{x,y} \sum_{i,i',j,j'=1}^d
 |L^{ii'}_{x,y,n,m}|^2=\\
 && d^2 \sum_{n,m}' \sum_{\beta}
 \frac{\lambda_{\beta+Dn}\lambda_{\beta+Dm}}{\sum_k
 \lambda_{\beta+Dk}}
 \le 2 d^2 \sum_\beta \sum_{n=1}^\infty \lambda_{\beta+Dn}\\
 && \le 2 d^2 \sum_{x=D}^\infty \lambda_x=2d^2\epsilon(D).
 \end{eqnarray*}

Now, it only remains to bound $\Delta t$. Again making use of the Cauchy--Schwartz inequality and the inequality
\[
 \left| 1- \frac{1}{1+x} \frac{1}{1+y} \right|
 \le x+y
 \]
 we get
\[\Delta t\le d^2\sum_{\alpha=1}^D\sqrt{\lambda_\alpha}\sum_{\beta=1}^D\sum_{n=1}^\infty
 \frac{\lambda_{\beta+Dn}}{\sqrt{\lambda_\beta}}.\]
The bound states in the lemma is given by $\Delta r+\Delta s+\Delta t$.\qed

The usefulness of this lemma relies in the fact that the spectrum $\{\lambda^i\}$ is typically decaying very fast
for noncritical systems, and hence the upper bound decays typically very fast with increasing $D$. Let us
illustrate this on the hand of e.g. the XXZ Hamiltonian on a chain
 \[
 \mathcal{H}=-\sum_{<i,j>}S^x_i S^x_j+S^y_iS^y_j+\Delta
 S^z_iS^z_j.
 \]
For $\Delta<-1$, the spectrum of the halve-infinite chain can be calculated exactly \cite{Peshel}. All
eigenvalues are of the form $c(z)z^n$  where
\begin{eqnarray*}
z&=&\exp\left(-2{\rm arcosh}(|\Delta|)\right)\\
1/c(z)&=&\sum_{n=0}^\infty a(n)z^n
\end{eqnarray*}
and where $a(n)$ denotes the degeneracy of each eigenvalue and is given by the number of possible partitions of n
into odd parts. The asymptotics of $a(n)$ are known exactly \cite{Sloane}:
\[ a(n)\sim \frac{1}{4}\frac{e^{\frac{\pi}{\sqrt{3}}\sqrt{n}}}{3^{1/4} n^{3/4}}.\]
To determine the power $n(D)$ of the $(D+1)^{th}$ largest eigenvalue $c(z)z^n(D)$ (taking into account degeneracy
off course), we have solve the equation
\[D=\sum_{m=0}^{n(D)} a(m)\]
for $n(D)$. Asymptotically, this becomes
\[ D\sim
\frac{3^{1/4}}{2\pi}\frac{e^{\frac{\pi}{\sqrt{3}}\sqrt{n(D)}}}{n(D)^{1/4}},\hspace{.5cm}n(D)\sim
\left(\frac{\log(D)}{\pi/\sqrt{3}}\right)^2.\]

$\epsilon(D)$, the sum of all eigenvalues smaller than the $D$ largest ones, can now readily be calculated as
\[\epsilon(D)=c(z)\sum_{m=n(D)}^{\infty}a(m)z^m\sim\exp\left(-\frac{3|\log z|}{\pi^2}\left(\log D\right)^2\right).\]
This last expression decays faster than any inverse power of $D$. A similar calculation holds for $\gamma_D$,
showing that the upper bound in lemma $3$ decays faster than any polynomial in $D$. This is a very nice
illustration for the superpolynomial accuracy of the infinite DMRG method for the XXZ-model. The other integrable
models will have a very similar behavior, and it is expected that non-integrable models share the same features.
Of course, the upper bounds derived here are not tight, and in practice the accuracy of a DMRG will be much
better than derived here.

\section{Conclusion}

In conclusion, we highlighted the importance of the concept of local reduced density operators of translational
invariant systems, their approximation using matrix product states, and the connection with frustration and
monogamy properties of entanglement.  We quantified how well matrix product states approximate exact ground
states of 1-D quantum spin systems as a function of the entropy of blocks of spins, and showed that the
complexity for representing ground states of 1-D systems (even critical ones) as MPS scales polynomially in the
number of spins.

We thank O. Sylju{\aa}sen for sending us his Monte Carlo data about the 2-D XXZ-model, G. Vidal for interesting
discussions and T. Osborne for pointing out the work \cite{datahiding}. Work supported by European projects, der
Bayerischen Staatsregierung and the Gordon and Betty Moore Foundation.


\begin{thebibliography}{99}

\bibitem{Wilson} K.G. Wilson, Rev. Mod. Phys. {\bf 47}, 773 (1975).

\bibitem{DMRG} S. R.  White, Phys. Rev. Lett.  {\bf 69}, 2863 (1992).

\bibitem{MPS2} S. Ostlund and S. Rommer, Phys. Rev. Lett. {\bf 75}, 3537 (1995); J. Dukelsky {\it et al.},
Europhys.Lett. {\bf 43}, 457 (1997).

\bibitem{Vwilson}  F. Verstraete, A. Weichselbaum, U. Schollwöck, J. I. Cirac, Jan von Delft, cond-mat/0504305.

\bibitem{Lloyd} S. Lloyd, Science {\bf 273} 1073 (1996).

\bibitem{Fahri} E. Fahri {\it et al}, quant-ph/0001106.

\bibitem{Murg} See, for example, V. Murg and J. I. Cirac,
Phys. Rev. A {\bf 69}, 042320 (2004).

\bibitem{CZ} J. I. Cirac and P. Zoller, Phys. Today, March,
38 (2004).

\bibitem{Werner} R.F. Werner, Lett. Math. Phys. 17, 359 (1989).

\bibitem{Terhal} B.M. Terhal, quant-ph/0307120 (2003); B.M. Terhal, A.C. Doherty, D. Schwab, Phys. Rev. Lett. 90, 157903 (2003).

\bibitem{Wolf}  M.M. Wolf, F. Verstraete, J.I. Cirac, Int. Journal of Quantum Information {\bf 1}, 465
(2003).

\bibitem{Doherty} M.R. Dowling, A.C. Doherty and S.D. Bartlett,
Phys. Rev. A {\bf 70}, 062113 (2004).


\bibitem{Werner2} R. F. Werner, Phys. Rev. A {\bf 40}, 4277
(1989).

\bibitem{Wolf0} M.M. Wolf, F. Verstraete, J.I. Cirac, Phys. Rev. Lett. {\bf 92},
087903 (2004).

\bibitem{Yang} C. N. Yang and C. P. Yang Phys. Rev. {\bf 150}, 321 (1966); Phys. Rev. {\bf 150}, 327
(1966).

\bibitem{2D} F. Verstraete and J.I. Cirac, cond-mat/0407066

\bibitem{Suljavson} O.F. Sylju{\aa}sen, Phys. Lett.  A {\bf 322}, 25 (2004).

\bibitem{Lin} H.-Q. Lin, J.S. Flynn and D.D. Betts, Phys. Rev. B
{\bf 64}, 214411 (2001).

\bibitem{Fannes} M. Fannes, B. Nachtergaele and R. F. Werner, Comm. Math. Phys.
{\bf 144}, 443 (1992); A. Kl\"umper, A. Schadschneider and J. Zittartz, J. Phys. A {\bf 24}, L955 (1991).

\bibitem{VPC04} F. Verstraete, D. Porras, and J. I. Cirac, Phys. Rev. Lett. {\bf 93}, 227205 (2004).

\bibitem{2Dd} F. Verstraete and J.I. Cirac, Phys. Rev. A.  {\bf 70},
060302(R) (2004).

\bibitem{Nishino} T. Nishino \emph{et al.}, Prog. Theor. Phys. {\bf
105}, 409 (2001).

\bibitem{DMRG1} U. Schollw\"ock, Rev. Mod. Phys. {\bf 77}, 259 (2005).

\bibitem{Hastings} M. B. Hastings, Phys. Rev. Lett. {\bf 93}, 140402 (2004).

\bibitem{datahiding} P. Hayden, D. Leung, P. W. Shor, A. Winter, Commun. Math.
Phys. {\bf 250}, 371 (2004).


\bibitem{VidalMPS} G. Vidal, Phys. Rev. Lett. {\bf 91}, 147902
(2003).

\bibitem{MPSrenorm} F. Verstraete, J.I. Cirac, J.I. Latorre, E. Rico, M.M. Wolf, Phys. Rev. Lett. {\bf 94}, 140601  (2005).

\bibitem{Peshel} I. Peschel, M. Kaulke and O. Legeza, Ann. Physik
(Leipzig) {\bf 8}, 153 (1999).

\bibitem{Kitaev} A. Kitaev, private communication.

\bibitem{Korepin} B.-Q. Jin and  V.E. Korepin,  J. Stat. Phys. {\bf 116}, 79
(2004).

\bibitem{Cardy} P. Calabrese and J. Cardy, J. Stat. Mech. (2004) P06002.


\bibitem{Vidal} G. Vidal, J. I. Latorre, E. Rico, and A. Kitaev, Phys. Rev. Lett. {\bf 90}, 227902 (2003).

\bibitem{Fannescompl} M. Fannes, B. Nachtergaele and R.F. Werner, Lett.
Math. Phys. {\bf 25}, 249 (1992).

\bibitem{VGC04} F. Verstraete, J. J. Garcia-Ripoll, J.I. Cirac, Phys.
Rev. Lett. {\bf 93}, 207204 (2004).

\bibitem{Sloane}  N. J. A. Sloane, The On-Line Encyclopedia of Integer Sequences, Sequence A000009.

\end{thebibliography}
\end{document}